\documentclass[twocolumn,secnumarabic,amssymb,nobibnotes,aps, prb,superscriptaddress]{revtex4-1}
\usepackage{graphicx}
\usepackage{subfigure}
\usepackage{mathrsfs}
\usepackage{amsthm}
\usepackage{amsmath}
\usepackage{float}
\usepackage{color}
\usepackage{natbib}
\begin{document}
\title{Numerical Study of Universal Conductance Fluctuation in Three-dimensional Topological Semimetals}
\author{Yayun Hu}
\affiliation{International Center for Quantum Materials, School of Physics, Peking University, Beijing 100871, China}
\author{Haiwen Liu}
\affiliation{Center for Advanced Quantum Studies, Department of Physics,
Beijing Normal University, Beijing 100875, China}
\author{Hua Jiang}
\email{jianghuaphy@suda.edu.cn}
\affiliation{College of Physics, Optoelectronics and Energy, Soochow University, Suzhou 215006, China}
\author{X. C. Xie}
\affiliation{International Center for Quantum Materials, School of Physics, Peking University, Beijing 100871, China}
\affiliation{Collaborative Innovation Center of Quantum Matter, Beijing 100871, China}

\begin{abstract}
We study the conductance fluctuation in topological semimetals. Through statistic distribution of energy levels of topological semimetals, we determine the dominant parameters of universal conductance fluctuation (UCF), i.e., the number of uncorrelated bands $k$, the level degeneracy $s$, and the symmetry parameter $\beta$. These parameters allow us to predict the zero-temperature intrinsic UCF of topological semimetals by the Altshuler-Lee-Stone theory. Then, we obtain numerically the conductance fluctuations  for topological semimetals of quasi-1D geometry. We find that for Dirac/Weyl semimetals, the theoretical prediction coincides with the numerical results. However, a non-universal conductance fluctuation behavior is found for topological nodal line semimetals, i.e., the conductance fluctuation amplitude increases with the enlargement of SOC strength. We find that such unexpected parameter-dependent phenomena of conductance fluctuation are related to Fermi surface shape of 3D topological semimetals. These results will help us to understand the existing and future experimental results of UCF in 3D topological semimetals.
\end{abstract}

\maketitle
%\tableofcontents
\section{Introduction}
Three-dimensional topological semimetals, including topological nodal point semimetals and topological nodal line semimetals, etc, are the celebrated paradigm of topological states\cite{RN541}. The Dirac/Weyl semimetals are examples of topological nodal point semimetals, which are characterized by the existence of linear touching nodes between conduction and valance bands in the bulk momentum space and Fermi-arc states on the surface. In comparison, for topological nodal line semimetals, conduction and valance cross at closed lines instead of discrete points\cite{RN542,RN538,RN539,RN537}. Following the theoretical proposals of topological semimetals in recent years\cite{RN458,RN461,RN459,RN460,RN467,RN468}, these novel materials have soon been experimentally realized\cite{RN462,RN463,RN469,RN470}. Due to the unique band structure, topological semimetals exhibit a number of exotic transport properties, e.g. negative magnetoresistance induced by chiral anomaly\cite{RN608,RN532,RN471}, high mobilities\cite{RN533,RN534}, electric-optic phenomena\cite{RN526,RN528}, which have been extensively explored. In addition to the quantum correction to conductance, the statistical properties of conductance in topological semimetals, e.g., the conductance fluctuation has also generated great interest. In particular, in the conductance fluctuation experiment of 3D Dirac semimetal Cd$_3$As$_2$, a $\frac{1}{2\sqrt{2}}$ suppression of conductance fluctuation in response to magnetic field has been reported\cite{RN351}.

Conductance fluctuation reflects the interference among different trajectories of electrons traversing inside the metal. According to the universal conductance fluctuation(UCF) theory\cite{RN4,RN3,RN486}, for an ensemble of diffusive metals, their zero temperature conductances follow a gaussian distribution with a fixed width $\Delta G$, which is a universal value of order $e^2/h$, determined only by two factors: (1) the symmetry and (2) the dimension of the system. As long as the diffusion condition (i.e., size L of the metal is much larger than the mean-free path $l$ and much smaller than the localization length $\xi$) is satisfied, $\Delta G$ is independent of other system details, such as the disorder strength $W$, the Fermi energy $E_F$, and the conductance $G$ itself.

The first factor that influences the conductance fluctuation is symmetry. According to the random matrix theory(RMT), there exist three types of classical ensembles: (1) the orthogonal ensemble, characterized by
symmetry index $\beta=1$ when both the time-reversal and spin-rotation symmetries are present; (2) the unitary ensemble $\beta=2$ if time-reversal symmetry is broken; and (3) symplectic ensemble $\beta=4$ if the spin-rotation symmetry is broken while time-reversal symmetry is preserved. The UCF amplitude for an isotropic rectangular material with transversal length $L_x, L_y$ and longitudinal length $L_z$ is
\begin{equation}\label{ucf}
  \Delta G=c_d\sqrt{\frac{ks^2}{\beta}},
\end{equation}
in the unit of $\frac{e^2}{h}$, where $c_d$ is a dimension-dependent constant which we will address later, $k$ is the number of independent energy bands involved in conduction, and $s$ is the symmetry-protected energy level degeneracy which is not removed by the random Hamiltonian\cite{RN486}.

\begin{figure}
\centering
\includegraphics[scale=0.6]{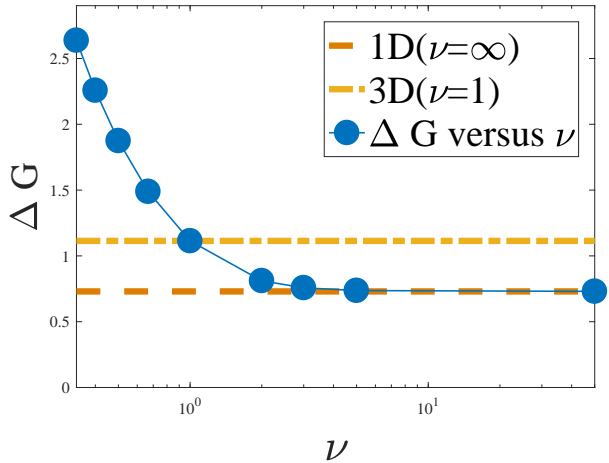}
\caption{$\Delta G$ versus $\nu=L_z/L_x=L_z/L_y$ with open boundary conditions. $\nu=1/3,2/5,1/2,2/3,1,2,3,5,50$ from left to right.}
\label{Fig1}
\end{figure}

Spatial dimension of the sample is the second deterministic factor of the conductance fluctuation amplitude. The coefficient $c_d$ in Eq.(\ref{ucf}) depends on the relative ratio between the lengths of the sample along three directions, i.e., on the normalized lengths $\nu_x=L_z/L_x$ and $\nu_y=L_z/L_y$, rather than the absolute lengths $L_{x,y,z}$. Fig.\ref{Fig1} shows the relation between $\Delta G$ versus size ratio $\nu=L_z/L_x=L_z/L_y$, which is calculated from Eq.(2.8) in Ref.[\onlinecite{RN4}]. $\Delta G$ quickly converges to 1D value as $\nu>5$.
In the literature, a quasi-1D material means $\nu_x=\nu_y\sim\infty$, quasi-2D corresponds to one of $\nu_{x,y}$ is unity while the other diverges, and 3D material is of cube shape with $\nu_{x,y}=1$. Here $c_d=0.365,0.43,0.55$ for dimension $d=1,2,3$, respectively. If periodic boundary condition(PBC) is applied along $L_x(L_y)$ direction, $c_{2,3}$ decreases by an amount because of removal of diffusion modes that can not exist under the PBC\cite{RN545,RN281}, while $c_{1}$ remains unchanged since in quasi-1D geometry, transversal diffusion modes do not contribute to conductance fluctuation with either open or periodic boundary condition.

In the study of topological insulators, the above properties of conductance fluctuation have been utilized to distinguish surface/bulk states transport in Bi$_{2}$Te$_{2}$Se microflakes\cite{RN492}, to manifest the edge-bulk states mixing in 2D HgTe quantum well\cite{RN489} and surface helical states mixing in 3D Bi$_{2}$Se$_{3}$\cite{RN520}, and to indicate various transitions between symmetry ensembles in quasi-1D Kane-Mele system\cite{RN493}. In 2D topological semimetal, graphene, the intrinsic conductance fluctuation has been investigated both analytically\cite{RN548,RN78} and numerically\cite{RN550,RN478}. In addition to spin degeneracy, survival of valley degeneracy under long-range disorder in graphene near the linear-dispersion point doubles the conductance fluctuation amplitude, which has been experimentally confirmed\cite{RN122}.

Unlike 2D graphene, there are no sublattice in 3D topological semimetals. However, the Fermion doubling theory requires Dirac points appear in pairs, it is thus tempting to ask whether or not valley degeneracy still hold in 3D topological semimetals when considering its conductance fluctuations? Also, though edge states in 2D graphene are spatially separate and do not contribute to conductance fluctuation in weak disorder, surface states in 3D topological semimetals will experience scattering due to impurity and hence contribute to conductance fluctuation\cite{RN482,RN494,RN710}. The existing theory for conductance fluctuation in 2D Dirac materials like graphene should be theoretically reexamined before applying to the 3D topological semimetals.

On the experimental side, nice data of finite-temperature conductance fluctuation of Cd$_3$As$_2$ at Fermi energy $E_F$ both close to and away from the Dirac node, has been reported\cite{RN351}. However, the complexity of the material, e.g., the existence of spin-orbit coupling (SOC) among multiple orbits, poses great challenges to the subsequent analysis of the experimental data. Besides, because of decoherence due to finite temperature and inhomogeneous vacancies in transport\cite{RN535}, there is still no experimental work on the intrinsic conductance fluctuation in 3D topological semimetals yet. Therefore, in order to provide a thorough understanding of the previous and future experiments, and to study the effects of SOC as well as to check the validity of existing theory, it is important to investigate the intrinsic conductance fluctuation of 3D topological semimetals.

In this article, we numerically study the universal conductance fluctuation in the presence of short-range disorder in 3D topological materials. We consider UCF for Fermi energy  both close and away from nodal point, as well as the influence of magnetic field and SOC strength. The UCF value consistent with random matrix theory is confirmed for Dirac/Weyl semimetals in a large range of parameters. However, for nodal line semimetals, an unexpected parameter-dependent behaviour of UCF emerges, i.e., the conductance fluctuation increases with an increase of SOC strength. Furthermore, we find that in addition to symmetry and dimension, the shape of the Fermi surface also has deterministic influence on UCF. The parameter-dependent UCF can be well explained by the SOC dependence of the Fermi surface shape of 3D topological semimetals.

This paper proceeds as follows: we firstly describe the model and method in section \uppercase\expandafter{\romannumeral2}. The nearest neighbour-distribution of energy levels is calculated to classify the symmetry of the Hamiltonian in section \uppercase\expandafter{\romannumeral3}. The numerical results of conductance fluctuation in Dirac/Weyl semimetals are presented in section \uppercase\expandafter{\romannumeral4}. The unusual non-universal conductance fluctuation for nodal line semimetals is discussed in section \uppercase\expandafter{\romannumeral5} followed by a brief conclusion in section \uppercase\expandafter{\romannumeral6}.
\section{Model and method}
Our numerical procedure closely follows Ref.[\onlinecite{RN481}]. We consider a topological semimetal placed in region $0<z<L_z$ with transverse size $L_x\times L_y$, described by the Hamiltonian:
\begin{equation}
  H=H_0+U(\textbf{r}),
\end{equation}
where $U(\textbf{r})$ is a random potential uniformly distributed on $[-W,W]$, and $H_0$ is the clean topological semimetal Hamiltonian described by\cite{RN480,RN541}
\begin{align}\label{socH0}
\setlength{\leftskip}{-12pt}
\mspace{-45mu} \nonumber &H_0(\textbf{k}) = \\
\mspace{-45mu} &\left(
   \begin{array}{cccc}
    M(\textbf{k}) & A\textbf{k}_++iB\textbf{k}_z &  D\textbf{k}_- & 0 \\
    A\textbf{k}_--iB\textbf{k}_z & -M(\textbf{k}) &  0 & 0 \\
    D\textbf{k}_+ & 0 &M(\textbf{k}) & -A\textbf{k}_-+iB\textbf{k}_z \\
    0  & 0 &  -A\textbf{k}_+-iB\textbf{k}_z &  -M(\textbf{k})
   \end{array}
  \right),
\end{align}
where $M(\textbf{k})=M_0-M_zk_{z}^{2}-M_xk_{x}^{2}-M_yk_{y}^{2}$, $k_{\pm}=k_x\pm ik_y$. $A$ and $B$ are the strength of SOC coupling inverted bands $\pm M(\textbf{k})$, while $D$ is the SOC strength coupling two $M(\textbf{k})$ orbits. Throughout the paper, we fix $M_0=-0.4, M_z=M_x=M_y=-0.5$ and we consider UCF for Dirac/Weyl semimetal($B=0$) and nodal line semimetal($B\neq0$ and $A=D=0$), respectively\cite{RN541}. The topological semimetal is connected to two infinite ideal leads at $z<0$ and $z>L_z$, which is also modeled by the clean Hamiltonian $H_0$.

We discretize $H$ on a cubic lattice, and the externally applied magnetic filed $\vec{B}$ can be considered through two effects. The first effect is the Zeeman term
\begin{eqnarray}\label{Hmz}%此处应该用直积形式来该写！！
\nonumber & &  H_{Zeeman} = \\
& &\left(
   \begin{array}{cccc}
    m_z & 0 & m_x-im_y & 0\\
    0 & m_z & 0 & m_x-im_y\\
    m_x+im_y & 0 & -m_z & 0 \\
    0 & m_x+im_y & 0 & -m_z \\
   \end{array}
  \right),
\end{eqnarray}
where the direction of $\vec{m}$ is parallel to $\vec{B}$ and its magnitude is proportional to $|\vec{B}|$. The other effect is the modification of hopping phase, e.g., for magnetic field applied along z direction, $t_x\rightarrow t_xe^{i\phi}$, where $\phi$ measures the magnetic flux through a unit lattice square.

We numerically compute the zero-temperature conductance using the Landauer-B\"{u}ttiker formula\cite{RN543} $G=\frac{e^2}{h}Tr[\Gamma_LG^r\Gamma_RG^a]$,  where $G^r(E_F)=[G^a(E_F)]^\dag=[E_F-H^{cen}-\Sigma_L-\Sigma_R]^{-1}$ is the retarded Green's function, $H^{cen}$ is the central region Hamiltonian, $\Gamma_{L/R}=i[\Sigma^r_{L/R}-\Sigma^a_{L/R}]$ is the line width function, $\Sigma_{L/R}$ is the self-energy of the left/right lead. The conductance fluctuation $\Delta G$ is calculated as the standard deviation of conductance $G$ for an ensemble of  disorder, $\Delta G=\langle(G-\overline{G})^2\rangle^{\frac{1}{2}}$, averaged over at least 400 ensembles.
\section{Symmetry Analysis for Dirac/Weyl semimetals}
\begin{figure}
\centering
\setlength{\leftskip}{-6pt}
\includegraphics[scale=0.43]{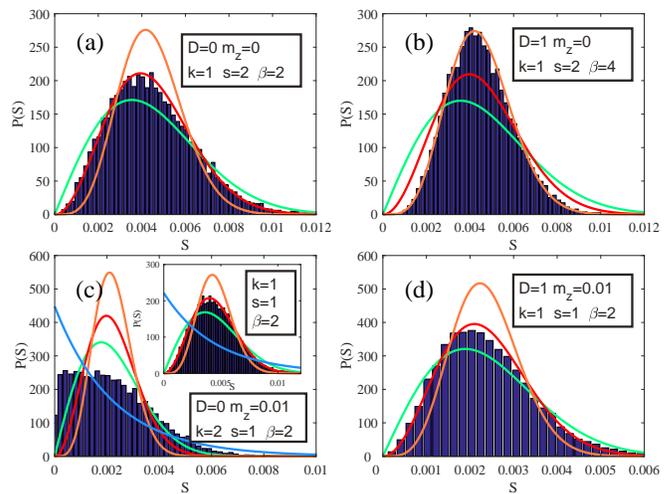}
\caption{Nearest neighbour energy level distance distribution $P(S)$ for different symmetry classes of Hamiltonian $H$, $L_x=L_y=L_z=10$, $A=0.5$, Fermi energy $E_F=0$, $W=7$, averaged over $1000$ ensembles. (a) $D=0$, $m_z=0$. (b) $D=1$, $m_z=0$. (c) $D=0$, $m_z=0.01$. (d) $D=1$, $m_z=0.01$. The solid green (red, orange) line represents the Wigner surmise of an orthogonal (unitary, symplectic) ensemble and blue line represents Poisson distribution, respectively.}
\label{Fig2}
\end{figure}
According to the UCF theory Eq.(\ref{ucf}), in order to theoretically predict the amplitude of conductance fluctuation, the indices $k$, $s$ and $\beta$ must be determined . We first count these symmetry indices directly by computing the energy level statistics\cite{RN12}. For disordered material of size $L_x=L_y=L_z=L$, we obtain the $4L^3$ eigenvalues sorted in descending order $E_i (i=1,2,\dots,4L^3)$ by exact diagonalization of $H$, and calculate the nearest neighbour distance $S_i=E_i-E_{i+1}(i=1,2,\dots,4L^3-1)$. The probability distribution of $S$ is defined as $P(S)=\langle\sum_i\delta(S-S_i)\rangle$, where the symmetry-protected degeneracy $S_i=0$ is neglected in the summation. Because of
energy repulsion in the metallic region, the Wigner surmise of probability distribution $P(S)$ of nearest neighbour distance $S$ reads, $P_{\beta}(S)=a_{\beta}(\frac{S}{\Delta})^{\beta}e^{-b_\beta (\frac{S}{\Delta})^2}$, where $\beta=1,2,4$, $\Delta$ is the average level distance of all $S_i$ under consideration, $a_{\beta}$ and $b_{\beta}$ are $\beta$ and $\Delta$ dependent constants that are given in Ref.[\onlinecite{RN12}].

In order to be consistent with recent experiments in Dirac semimetals\cite{RN351}, we will firstly set $B=0$. We now consider the symmetry transition of the system driven by external magnetic field and SOC strength $D$. The magnetic field is considered through Zeeman energy $m_z$. It is numerically confirmed that both Zeeman energy added along x/y direction and magnetic field entered through addition of hopping phase influences the symmetry transition of the system in the same way, i.e., by breaking of time-reversal symmetry.

For $D=0$ and in the absence of a magnetic field, we find an exact $s=2$ degeneracy from level statistics. This is because, though the upper block and lower blocks (which we refer to as $H_\uparrow$ and $H_\downarrow$ respectively in the following) decouple in this case, they are still related by time-reversal symmetry, $H_\uparrow=T^{-1} H_\downarrow T$, therefore giving an exact copy of energy levels (Kramer counterpart). Furthermore, by comparing with the fitting curves of different symmetries, it is clear that $P(S)$ at $E=0$ coincides well with $P_{2}(S)$ [see Fig. \ref{Fig2}(a)], with subscript $\beta=2$ indicating a unitary ensemble. Thus we find both $H_\uparrow$ and $H_\downarrow$ belong to unitary ensemble with indices $k=1,\beta=2,s=1$. Since they are degenerate counterparts, the corresponding indices for $H$ are $k=1,\beta=2,s=2$. We note that the unitary limit has already been reached for our sample size $L_x=L_y=L_z=10$, which is studied in the above computation. With the same method, we also determine $k,s$ and $\beta$ for the following cases.

When SOC strength $D$ is turned on, the two blocks merge and the system manifests symplectic statistics with $k=1,\beta=4,s=2$ [see Fig. \ref{Fig2}(b)]. When magnetic field further comes into play by adding a Zeeman energy $m_z=0.01$, time reversal symmetry is broken and we find the system is unitary with $\beta=2$ [see Fig. \ref{Fig2}(d)]. When $D$ is set to zero and Zeeman term holds, distribution of $P(S)$ does not satisfy any ensemble fitting [see Fig. \ref{Fig2}(c)]. The system is still unitary but the upper and lower blocks contribute independently to the energy levels of $H$, leading to suppression of energy level repulsion. This is witnessed through the peak around $S=0$ in $P(S)$ [see Fig. \ref{Fig2}(c)], in contrast to the vanishing $P(S)$ at $S=0$ as a result of avoided level crossing [see Fig. \ref{Fig2}(a,b,d)]. If we focus only on $P(S)$ of either $H_\uparrow$ or $H_\uparrow$, as shown in the inset of Fig. \ref{Fig2}(c), indices $k=2,\beta=2,s=1$ are clearly confirmed.

The nearest neighbour distribution for long-range disorder is also investigated, and the previous conclusions still hold. Also, $P(S)$ for energy far away from nodal point gives the same symmetry indices. Thus we conclude that the valley degeneracy need not be considered in 3D topological semimetals. Furthermore, the symmetry class of 3D topological semimetals is independent of either energy or disorder type.
\section{Conductance fluctuation in Dirac/Weyl semiemtals}
\begin{figure}
\centering
\setlength{\leftskip}{-12pt}
\includegraphics[scale=0.47]{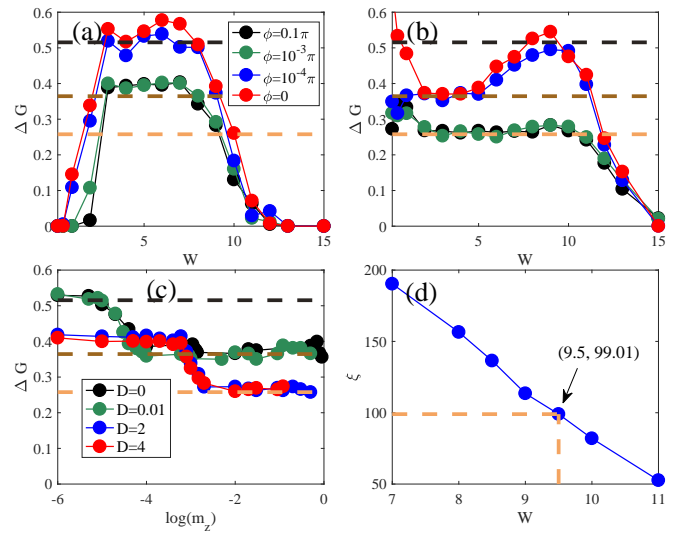}
\caption{Evolution of UCF with symmetry parameters, $L_x=10,L_y=20,L_z=100$, $A=1$, $E_F=0.01$. (a) $\Delta G$ versus $W$, in the condition of $D=0$, $\phi=0\sim0.1\pi$. (b) $\Delta G$ versus $W$, in the condition of $D=3$, $\phi=0\sim0.1\pi$. The same legend in (a) applies to (b). (c) UCF for $D=0\sim4$, $m_z=10^{-6}\sim 1$. Horizontal lines in panels (a), (b) and (c) from up to down correspond to $\Delta G=c_1/\sqrt{2}, 2, 2\sqrt{2}=0.516, 0.365, 0.258$. (d) Localization length $\xi$ versus disorder strength $W$, calculated by transfer matrix method with $L_x=10,L_y=20$ and periodic boundary conditions are applied along each transverse dimension.}
\label{Fig3}
\end{figure}
It is predicted that the zero temperature conductance fluctuation only depends on the relative ratios between lengths along three directions $L_x,L_y,L_z$ of the material with a given symmetry\cite{RN4}. Our numerical calculation shows that if transverse lengths is small, for example, $L_x=5,L_y=5,L_z=500$, the conductance fluctuations will have a strong fermi surface dependence and the statistics of energy levels $P(S)$ will deviate from the anticipated Wigner-Dyson surmise. This means the ergodicity hypothesis is not satisfied in small systems. Only for large enough transverse dimensions (i.e., $L_x,L_y>8$) can one get rid of the finite size effect. For a dimension of $L_x=10,L_y=20,L_z=100$, the conductance fluctuation $\Delta G$ versus disorder strength W is calculated and is shown below under various parameters of $m_z$ and $D$.

When $D$ is set to zero and $\phi=0$ [see Fig. \ref{Fig3}(a)], $\Delta G$ consists of the same contribution from time reversal pairs of the upper and lower blocks of $H_{\uparrow}$ and $H_{\downarrow}$. With the onset of disorder, the conductance fluctuation increases from zero and then forms a plateau. Before formation of the plateau ($W<3$), the material is in ballistic transport region where conductance fluctuation has no universal value (i.e., disorder depedent). For disorder strength within $3<W<8$, conductance fluctuation is around $0.48\sim0.57$, showing very weak disorder strength dependence, which is a signature predicted by Altshuler-Lee-Stone theory in diffusive metals and has been used to numerically identify the region for universal conductance fluctuation. Plugging $k=1,\beta=2,s=2$ obtained from Fig. \ref{Fig2}(a) into Eq.(\ref{ucf}), we have the expected $\Delta G=0.365\sqrt{\frac{1\times2^2}{2}}=0.516$, which coincides with the plateau value. Here $\Delta G=0.516$ doubles the conductance fluctuation of $H_\uparrow$ or $H_\downarrow$, both of which belong to the 1D unitary metal case with $\Delta G=0.258$. We emphasize that index $k\neq2$ since $H_\uparrow$ and $H_\downarrow$ are strongly correlated rather than independent of each other, and the double in conductance fluctuation comes purely from the energy level degeneracy protected by time reversal symmetry. By further increasing disorder strength, the material becomes Anderson insulator with vanishing conductance as well as vanishing fluctuation. In comparison, when $D$ is zero while $\phi$ is greater than a certain threshold value $\phi_c$ [see Fig. \ref{Fig3}(a)], time-reversal symmetry is fully broken. The Kramer degeneracy is lifted, and $H_{\uparrow/\downarrow}$ contribute independently to $\Delta G$. The UCF decreases to $\Delta G\sim0.365$, consistent with Eq.(\ref{ucf}) for indices $k=2,\beta=2,s=1$ [see Fig. \ref{Fig2}(c)].

The result for strong SOC strength $D=3$ is shown in Fig. \ref{Fig3}(b). Without magnetic field, the system has conductance fluctuation $\Delta G\sim0.365$ between $W=2$ and $W=5$, consistent with symmetry indices $k=1,\beta=4,s=2$ for quasi-1D transport [see Fig. \ref{Fig2}(b)]. A second conductance fluctuation plateau [see Fig. \ref{Fig3}(b)] appears between $W=8$ and $W=10$, which is beyond the Altshuler-Lee-Stone theory. Similar second-plateau behaviour has been reported in Refs. [\onlinecite{RN536,RN545}], where the authors find the occurrence of this second plateau is related to metal-insulator transition in 2D unitary ($\beta=2$) and symplectic ($\beta=4$) systems. Adopting transfer matrix method\cite{RN527}, we calculate the localization length $\xi$ along z direction. It is found that at disorder strength $W=9.5$ where peak of the second plateau is located, $\xi=99.01$ coincides with the longitudinal length $L_z=100$ of the sample [see Fig. \ref{Fig3}(d)]. The same property is reported in Ref.[\onlinecite{RN536}]. We confirm the existence of such second fluctuation plateau in 3D topological semimetal system for large $D$.

Another feature is the anomalously large conductance fluctuation at small disorder strength ($W<1$) [see Fig. \ref{Fig3}(b)]. Similar behaviour is also observed in the numerical calculation of single-layered graphene, when long-range disorder is introduced\cite{RN550}. However, the conductance fluctuation anomaly therein disappears for short-range disorder, which is different from the results in our case. Besides, the results in graphene is explained as a finite-size effect according to a careful numerical examination\cite{RN478}. A possible explanation for this large conductance fluctuation phenomena is that, the system is still in ballistic transport region. Thus, the diffusion condition for the validity of UCF theory $l<L_{x,y,z}$ is not satisfied, where the mean free path $l$ is typically large for Dirac/Weyl fermions.

In the presence of a very weak magnetic field, $\phi=0$ to $\phi=10^{-4}\pi$, the above results remain unchanged [see lines in Fig. \ref{Fig3}(b)]. When the magnetic field increases beyond a certain threshold $\phi_c$, UCF value finally suppresses by a factor of $\sqrt{2}$ (decrease from the middle horizontal line to the bottom horizontal line), with new symmetry indices $k=1,\beta=2,s=1$ due to lifting of Kramer degeneracy. Moreover, the second conductance fluctuation plateau for $\beta=2$ is less stressed than that in the $\beta=4$ case, which is similar to 2D normal metals\cite{RN536}.

The evolution of universal conductance fluctuation $\Delta G$ with respect to magnetic field $m_z$ and SOC strength $D$ is shown in Fig. \ref{Fig3}(c). For each choice of $D$ and $m_z$, we plot $\Delta G$ versus $W$ as we do in Fig. \ref{Fig3}(a) and (b), to determine UCF amplitude. For small $D$ (i.e., $D=0,0.01$) and $m_z$ (i.e., $m_z<10^{-4}$), the highest plateau in Fig. \ref{Fig3}(c) corresponds to unitary ensemble with symmetry indices $k=1,\beta=2,s=2$. The conductance plateau decreases by a factor of $\frac{1}{\sqrt{2}}$ with increasing Zeeman energy $m_z$, and the system is in new symmetry class with indices $k=2,\beta=2,s=1$. For large $D=2,4$, the conductance fluctuation plateau emerges with symmetry indices $k=1,\beta=4,s=2$ at weak $m_z$. The height of the plateau also decreases by a factor of $\frac{1}{\sqrt{2}}$ with increasing Zeeman energy $m_z$. The new symmetry indices are, however, $k=1,\beta=2,s=1$ in this case. The transition between different plateaus indicates the transition between different symmetry classes. To summarize, coincidence between the height of these plateaus and the values predicted by Eq.(\ref{ucf}) manifests the applicability of the random matrix theory to a large range of parameters in Dirac/Weyl semimetals.
\section{Conductance fluctuation in topological nodal line semimetals}
\begin{figure}
\centering
\setlength{\leftskip}{-12pt}
\includegraphics[scale=0.44]{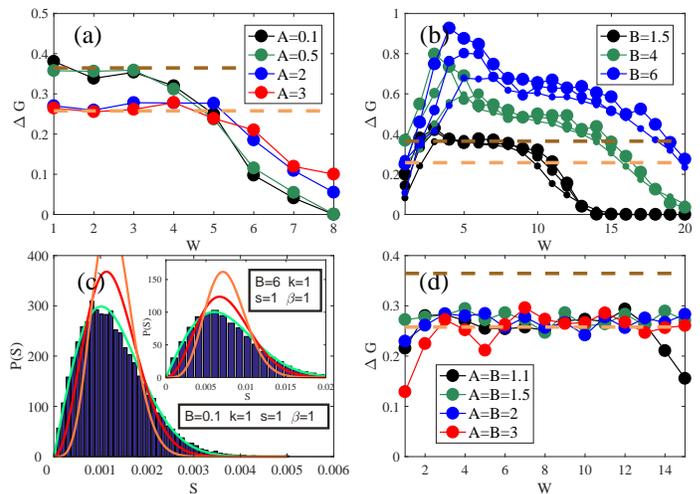}
\caption{(a)(b) $\Delta G$ versus $W$, in the condition of (a) size $L_x=15, L_y=15, L_z=100$, $B=0$, $E_F=3$, and (b) size $L_x=10, L_y=10, L_z=67$, ($L_x=15, L_y=15, L_z=100$, $L_x=20, L_y=20, L_z=133$), $A=0$, $E_F=0.5$. Larger marker size corresponds to larger sample size. (c) Nearest neighbour distribution $P(s)$ under $E_F=0.5$, $W=7$, $L_x=L_y=L_z=10$, $B=0.5, B=6(inset)$. (d) $\Delta G$ versus $W$, in the condition of $A=B$, $E_F=0.5$ and size $L_x=15, L_y=15, L_z=100$. In panels (a),(b) and (d), horizontal dashed lines correspond to $\Delta G=0.729/2, 2\sqrt{2}=0.365, 0.258$ respectively.}
\label{Fig4}
\end{figure}
In this section, we consider the effect of SOC strength $B$ on UCF of nodal line semimetals. We find a non-universal $B$-dependent conductance fluctuation arises. In order to demonstrate the result more clearly, we focus on the minimal topological semimetal Hamiltonian $H_\uparrow$ and show the numerical results in Fig. \ref{Fig4}.

Firstly, the $A$-dependence of $H_\uparrow$ is examined [see Fig. \ref{Fig4}(a)]. For small $A$, the two inverted bands are decoupled. Each band is labelled with orthogonal symmetry indices $k=1,\beta=1,s=1$, and thus each contributes a conductance fluctuation of amplitude $0.365$. However, when setting $E_F=0.5$, only the upward-open band is involved in conduction, therefore $\Delta G=0.365$. With the increase of $A$, the two inverted bands merge and the system now belongs to unitary symmetry class with indices $k=1,\beta=2,s=1$, and $\Delta G=0.258$, decreased by a factor of $\frac{1}{\sqrt{2}}$.

In the absence of $A$, $\Delta G=0.365$ for small $B (0.1<B<2)$, which is similar to the case of small $A$. However, with the increase of $B$ ($i.e., B>3$) [see Fig. \ref{Fig4}(b)], $\Delta G$ also increases monotonically, which is contrary to the deceasing of $\Delta G$ with increasing $A$. Conductance fluctuation plateau in this case is much larger than the expected amplitude in quasi-1D geometry. Does this anomalously large conductance fluctuation result from finite size effect? We fix the relative ratio between the lengths along the three dimensions of the material, $\nu_x=\nu_y=20/3$, and compare the numerical results for sample size $L_x=10, 15, 20$. It is found that $\Delta G$ does not decrease with increasing size, which means this non-universal fluctuation is not a finite-size effect. Then is the fluctuation enhancement caused by the change of the symmetry class? By examining nearest-neighbour energy level distance distribution $P(S)$ for both small and large SOC strength $B=0.1, 6$ [see Fig. \ref{Fig4}(c)], we confirm the system belongs to orthogonal ensemble, regardless of the strength of SOC. Therefore we conclude that such conductance fluctuation anomaly does not originate from the change of symmetry indices, either.

Furthermore, we test the response of the fluctuation from open to periodic boundary conditions along x and y directions. For small $B$, the fluctuation amplitude is insensitive to boundary conditions, which is a typical characteristic of quasi-1D material\cite{RN545,RN281}. On the contrary, for large $B$, where $\Delta G$ takes non-universal values, the conductance fluctuation plateau decreases for periodic boundary conditions. It is the typical of 2D and 3D materials but not for 1D material. This is not consistent with existing theory since we are clearly using quasi-1D geometry in our calculation.

The reasons for such inconsistency with existing theory are hidden in the Fermi surface shape in momentum space. To our knowledge, the existing UCF theory considers only isotropic band structure \cite{RN4,RN3,RN486}, which is also true in graphene with the Dirac cone dispersion\cite{RN548,RN78,RN550,RN478}. However, if one considers anisotropic band structure, one will find Fermi surface shape is another factor that plays a key role in conductance fluctuation, which has long been neglected in literature. The UCF has a similar dependence on Fermi surface shape in momentum space to the dependence on dimension size as plotted in Fig. \ref{Fig1}. That is, the larger the ratio of transversal Fermi surface size to longitudinal Fermi surface size, the larger the conductance fluctuation [see Appendix]. In the topological semimetals, SOC not only influences the symmetry of the material, but also gives rise to anisotropy in the band structure. In order to see the effect of SOC on the Fermi surface, one can rewrite $H_\uparrow$ in the discretized model as:
\begin{subequations}\label{Hsink2dB1}
\begin{align}
\label{Hsink2dB2}  H_\uparrow(\mathbf{k}) &= \mathcal{M}(\textbf{k})\sigma_z    +\frac{A}{a}[\sin(k_xa)\sigma_x-\sin(k_yb)\sigma_y] \\
\nonumber &+\frac{B}{c}\sin(k_zc)\sigma_y\,,  \\
\label{Hsink2dB3} \mathcal{M}(\textbf{k}) &=M_0-\frac{2M_z}{c^2}[1-\cos(k_zc)] \\
\nonumber &-\frac{2M_x}{a^2}[1-\cos (k_xa)]-\frac{2M_y}{b^2}[1-\cos(k_yb)]\,,
\end{align}
\end{subequations}
where we set $a=b=c=1$ in the numerical calculation. For Weyl semimetals $(B=0)$, the Fermi surface in momentum space satisfies $\cos(k_z)=[Q_1\pm \sqrt{E_F-A^2(\sin^2(k_x)+\sin^2(k_y))}]/(-2M_z)$, with $Q_1=-2M_z-2M_x-2M_y+M_0+2M_x\cos(k_x)+2M_y\cos(k_y)$. For nodal line semimetals $(A=0)$, the Fermi surface satisfies $\cos(k_z)=(-2M_zQ_1+\sqrt{Q_2})/(4M_z^2-B^2)$, with $Q_2=4M_z^2Q_1^2-(4M_z^2-B^2)(Q_1^2+B^2-E_F^2)$.

\begin{figure}
\centering
\setlength{\leftskip}{-18pt}
\includegraphics[scale=0.515]{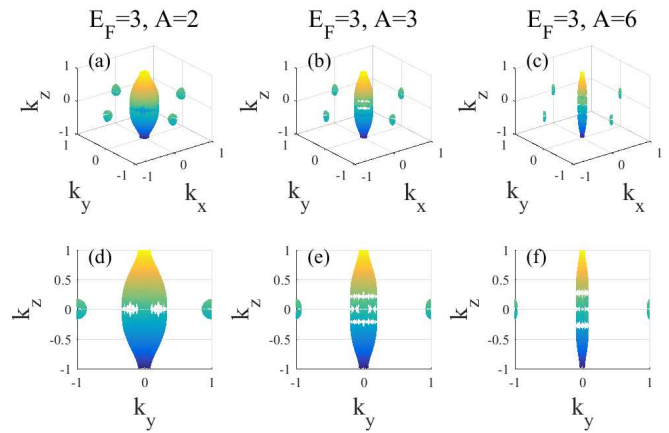}
\caption{Shapes of Fermi surface(upper panel) and the projections(lower panel) onto $k_y-k_z$ plane for $A=2,3,6$, $E_F=3, B=0$. Other parameters are the same as that used in  Fig. \ref{Fig4}(a).}
\label{Fig5}
\end{figure}
\begin{figure}
\centering
\setlength{\leftskip}{-18pt}
\includegraphics[scale=0.515]{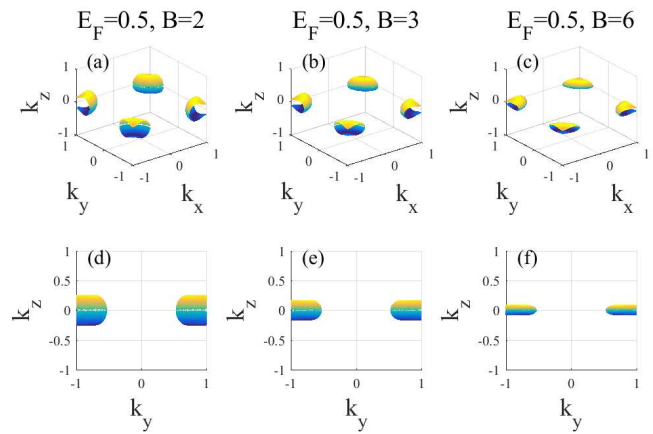}
\caption{Shapes of Fermi surface(upper panel) and the projections(lower panel) onto $k_y-k_z$ plane for $B=2,3,6$, $E_F=0.5, A=0$. Other parameters are the same as that used in  Fig. \ref{Fig4}(b).}
\label{Fig6}
\end{figure}

The Fermi surface and its corresponding projection onto $k_y-k_z$ plane in momentum space for varying SOC strength in Weyl semimetals and nodal line semimetals are shown in Fig. \ref{Fig5} and Fig. \ref{Fig6}, respectively. With the increase of $A$, the $E_F=3$ Fermi surface is stretched along the $k_z$ direction [see Fig. \ref{Fig5}(a)-(c)], which makes the whole Fermi surface look more like a quasi-1D shape in momentum space. This is demonstrated more clearly when the Fermi surfaces are projected onto the  $k_y-k_z$ plane [see Fig. \ref{Fig5}(d)-(f)]. The corresponding UCF under these parameters keep the quasi-1D value [see Fig. \ref{Fig4}(a)]. In contrast, with the increase of $B$, the Fermi surface is compressed along $k_z$ direction [see Fig. \ref{Fig6}(a)-(f)]. The corresponding UCF further increases with the further stretching of Fermi surface [see Fig. \ref{Fig4}(b)]. This is similar to increase of UCF with increasing size ratio $L_x(L_y)/L_z$. In comparison, when both longitudinal and transversal SOC are present, the Fermi surface will suffer less anisotropy and the UCF plateau will remain the quasi-1D value as expected [see Fig. \ref{Fig4}(d)].

An intuitive understanding of the Fermi surface dependence of UCF is as follows. The various diffusion modes contribute to UCF by certain functions of their eigenvalues $\lambda_{n_x,n_y,n_z}$\cite{RN3,RN4}, whose form have been generalized to anisotropic systems as Eq.(\ref{eigenvalue}) [see Appendix for detailed definitions]. The key thing to notice is the diffusion time $t_{\alpha}=\frac{L_{\alpha}^2}{D_{\alpha}}$ that governs the competition among diffusion modes along longitudinal and transversal directions. That is, the smaller the time ratio $t_{z}/t_{x,y}$ (the time for an electron to diffuse through length $L_z$ over the time through $L_{x,y}$\cite{RN544}), the larger the contribution to the UCF through summation of $n_{\alpha}$\cite{RN4}. For example [see Fig. \ref{Fig6}(a)-(f)], when Fermi surface is compressed along $\textbf{k}_{z}$ for a given $E_F$, the diffusion constant  $D_{z}=v_z^2\tau=\frac{4E_F^2\tau}{\hbar^2k_z^2}$ increases\cite{RN540}. As a result, the electron diffuses a shorter time $t_z$ than $t_{x,y}$, which enhances the transversal modes contribution to UCF [see Fig. \ref{Fig4}(b)]. Though our numerical result qualitatively shows that the evolutions of UCF are consistent with evolutions of Fermi surface shape, deep understanding of the Fermi surface shape's influence on the UCF still needs further study, which is beyond the scope of this work.
\section{Conclusions and discussions}
In summary, we have numerically studied the universal conductance fluctuation in 3D topological semimetals for both transversal and longitudinal SOC. We find the UCF plateaus are consistent with the RMT theory in a large range of parameters, which shows the applicability of UCF theory in Dirac/Weyl semimetals in most cases. However, a parameter-dependence of conductance fluctuation is found for nodal line semimetals, which can not be explained by previous theory. The origin of this anomaly can be understood as the dependence of Fermi surface shape on SOC strength. Compared with normal metals\cite{RN536}, where SOC only works on the symmetry of the material, in topological semimetals, SOC can also tune UCF by changing the band structure in addition to the symmetry.

By energy level statistics, we also determine that the valley degeneracy is irrelevant to the UCF in topological semimetals. In a recent experiment\cite{RN351}, the conductance fluctuation in Cd$_3$As$_2$ is shown to decay by a factor of $2\sqrt{2}$ when a magnetic field is applied. But according to our results for Dirac semimetal, the UCF amplitude should decay only by a factor of $\sqrt{2}$ [see Fig. \ref{Fig3}(a-b)]. Noticing that the authors assume valley degeneracy of Dirac semimetals in explaining the experimental results. However, our numerical result shows valley degeneracy should be lifted in the presence of disorder. We suggest that the $2\sqrt{2}$ decay may be caused by other factors, i.e., the magnetic field induced gap, the decoherence, etc., which still demands further investigation.
\section{Acknowledgement}
We thank Qing-Feng Sun, Yan-Xia Xing, Qiong Zhu, Chui-Zhen Chen for helpful discussion. One of us (Yayun Hu) would like to thank Juntao Song and Yijia Wu for their careful reading of the manuscript. This work is financially supported by NBRPC (Grants No. 2015CB921102 and No. 2014CB920901), NSFC (Grants No. 11534001, No. 11374219, No. 11504008, and No. 11674028), and NSF of Jiangsu Province, China (Grant No. BK20160007).
\begin{figure*}
\centering
\setlength{\leftskip}{-70pt}
\includegraphics[width=22cm, clip]{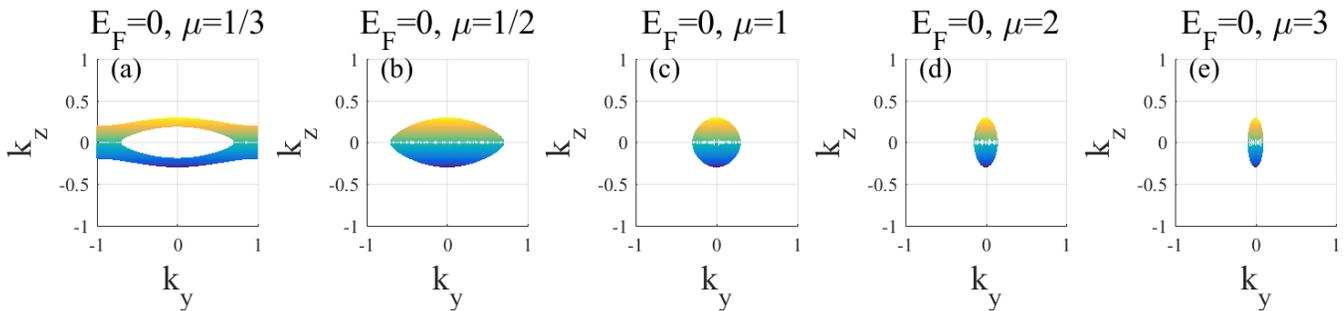}
\caption{Evolution of Fermi surface versus $\mu$, with $E_F=0$ fixed.}
\label{Fig7}
\end{figure*}%\includegraphics [width=15cm, viewport=11 220 1110 755, clip]{figure3.pdf}
\begin{figure}
\centering
\setlength{\leftskip}{-20pt}
\includegraphics[scale=0.46]{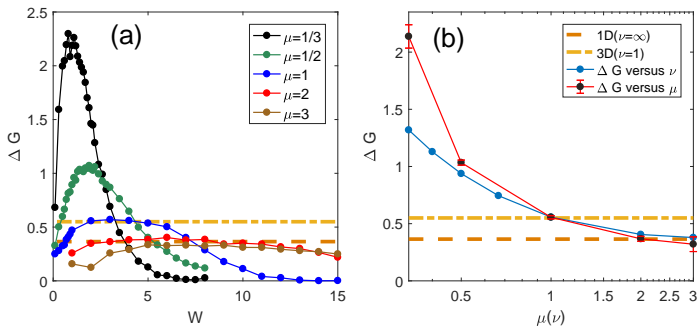}
\caption{(a)Evolution of conductance fluctuation versus disorder strength for different choice of $\mu$, other parameters are $E_F=0, L_x=L_y=L_z=10$. (b) Comparison between UCF amplitude versus Fermi surface anisotropy $\mu$ and spatial size anisotropy $\nu$. The error bars are estimated from $\Delta G$ at the plateau region in (a). The horizontal lines in both (a) and (b) are $\Delta G=0.55, 0.365$, corresponding to 3D and quasi-1D conductance fluctuation respectively.}
\label{Fig8}
\end{figure}

\section{Appendix}
We consider UCF for an anisotropic quadratic band structure caused by anisotropic effective mass, which can be described by Hamiltonian $M(\textbf{k})$ defined in Eq.(\ref{socH0}). For simplicity, we have neglected the degree of freedom of both spin and orbit. After discretization of $M(\textbf{k})$ into the lattice form, the effective Hamiltonian is $\mathcal{M}(\textbf{k})$, as defined in Eq.(\ref{Hsink2dB3}). Near the band bottom, $\mathcal{M}(\textbf{k})$ approximately reads $\mathcal{M}(\textbf{k})\approx M(\textbf{k})=M_0-M_zk_z^2-M_xk_x^2-M_yk_y^2$.

Throughout this appendix, we fix $M_0=-0.4, M_z=-0.5$, while tune $M_x=M_y$ to test the effect of Fermi surface shape on UCF. We use $\mu=\mu_{x,y}=\sqrt{|M_{x,y}/M_z|}$ to describe the degree of anisotropy in Fermi surface shape. The influence of Fermi surface shape on UCF amplitude is shown in Fig. \ref{Fig8}. For isotropic spherical Fermi surface ($\mu=1$), conductance fluctuation is consistent with 3D value of $\Delta G=0.55$ [see Fig. \ref{Fig8}(a)]. When $\mu=\frac{1}{2}<1$, the Fermi surface is an ellipsoid with principle axis placed in $k_x-k_y$ plane in momentum space [see Fig. \ref{Fig7}(b)]. With the decreasing of $\mu$, the transversal size ($k_{x/y}$) of Fermi surface is even larger than longitudinal ($k_z$) size in momentum space, and the corresponding conductance fluctuation also increases [see Fig. \ref{Fig8}(a)]. This property is similar to the enhancement of the UCF by increasing transversal size $L_{x(y)}$. In contrast, when $\mu>1$, the Fermi surface is an ellipsoid stretched along $k_z$ direction [see Fig. \ref{Fig7}(d-e)]. With the increase of $\mu$, UCF quickly decreases to quasi-1D amplitude $\Delta G=0.365$ [see Fig. \ref{Fig8}(a)]. Similarly, with the increasing of longitudinal size $L_z$, the decrease of UCF into $\Delta G=0.365$ is also expected.

The effect of Fermi surface shape on UCF is summarized and compared with the effect of spatial size in Fig.\ref{Fig8} (b). When Fermi surface or spatial size is stretched transversally(longitudinally), UCF increases(decreases). The similarity between the dependence of UCF on Fermi surface shape and spatial size can be understood as follows. To generalize the previous theory into anisotropic systems with Hamiltonian $\mathcal{M}(\textbf{k})$, one should substitute $D\nabla^2$ in Eq.(A8) with $D_x\nabla_x^2+D_y\nabla_y^2+D_z\nabla_z^2$ in Ref.[\onlinecite{RN4}], where $D_{\alpha}$, $\alpha=x,y,z$, are the anisotropic diffusion constants. This substitution changes the eigenvalues in Eq.(A10) into
\begin{equation}\label{eigenvalue}
  \lambda_{n}=\tau D_z(\frac{\pi}{L_z})^2[n_{z}^2+n_x^2\frac{D_x}{D_z}\frac{L_z^2}{L_x^2}
+n_y^2\frac{D_y}{D_z}\frac{L_z^2}{L_y^2}]
\end{equation}
where subscripts $n_{\alpha}$ are integers labeling the eigenvalues, $\tau$ is the scattering rate by impurities. And we have neglected the energy difference $\Delta E$ and the inverse inelastic scattering rate $\tau_{in}^{-1}$ since we are considering intrinsic conductance fluctuation. The appearance of coefficients in Eq.(\ref{eigenvalue}) $\frac{t_z^2}{t_{\alpha}^2}=\frac{D_{\alpha}L_z^2}{D_zL_{\alpha}^2}
=\frac{M_{\alpha}L_z^2}{M_zL_{\alpha}^2}=\mu^2_{\alpha}\nu^2_{\alpha}$
(where we have used the relation $\frac{D_{\alpha}}{D_{\beta}}=\frac{M_{\alpha}}{M_{\beta}}$\cite{RN540} and $t_{\alpha}=\frac{L_{\alpha}^2}{D_{\alpha}}$ is the characteristic time for an electron to diffuse through length $L_{\alpha}$\cite{RN544}), indicates that $M_{\alpha}$ and $L_{\alpha}$ can change the eigenvalues $\lambda_{n_x,n_y,n_z}$ equivalently, which next determine the UCF of the system. Thus one will not be able to determine the intrinsic conductance fluctuation of a metal merely by its size and symmetry. The Fermi surface shape information is also indispensable.
\bibliographystyle{apsrev4-1}
\bibliography{myreference1}
\end{document}